\documentstyle[12pt]{article}

\def\beqa{\begin{eqnarray}}
\def\eeqa{\end{eqnarray}}
\def\beq{\begin{equation}}
\def\eeq{\end{equation}}

\def\ad{\dot{a}}
\def\add{\ddot{a}}

\let\gam=\gamma
\let\alp=\alpha

\def\ad{\dot{a}}
\def\add{\ddot{a}}
\def\vol{\int d^4x\,\sqrt{-g}} 

\def\half{\frac{1}{2}}
\def\gu{g^{\mu\nu}}
\def\gd{g_{\mu\nu}}

\def\umunu{^{\mu\nu}}
\def\dmunu{_{\mu\nu}}

\def\dab{_{\alpha\beta}}

\def\ddemunu{_{;\mu\nu}}
\def\udemunu{^{;\mu\nu}}
\def\ddemu{_{;\mu}}  
\def\ddenu{_{;\nu}}  
\def\ddea{_{;\alpha}}  \def\udea{^{;\alpha}}

\def\pa{\partial}

\def\jmp{{\it J. Math. Phys.}\ }
\def\pr{{\it Phys. Rev.}\ }
\def\prl{{\it Phys. Rev. Lett.}\ }
\def\pl{{\it Phys. Lett.}\ }
\def\np{{\it Nucl. Phys.}\ }

\def\ijmp{{\it Int. Journ. Mod. Phys.}\ }

\def\cqg{{\it Class. Quantum Grav.}\ }

\def\grg{{\it Gen. Relativ. Grav.}\ }

\def\ncim{{\it Il Nuovo Cim.}\ }

\def\rmp{{\it Rev. Mod. Phys.}\ }

\def\ie{{\it i.e. }}
\def\eg{{\it e.g. }}
\def\jmp{{\it J. Math. Phys.}\ }
\def\pr{{\it Phys. Rev.}\ }
\def\pl{{\it Phys. Lett.}\ }

\def\rmp{{\it Rev. Mod. Phys.}\ }

\def\f{f(\phi)}
\def\fp{f'(\phi)}

\def\p{\phi}

\def\v{V(\phi)}
\def\vp{V'(\phi)}
\def\l{\cal L}
\def\g{\cal G}
\def\fo{\cal F}

\def\il{\item\label}

\begin{document}
\def\bib#1{[{\ref{#1}}]}
\begin{titlepage}

\title{Recovering the effective cosmological constant in 
       extended gravity theories}

 \author{{S. Capozziello$^{1,3}$, R. de Ritis$^{2,3}$, A.A. Marino$^{3,4}$}\\
 {\em {\small $^{1}$Dipartimento di Scienze Fisiche, "E.R. Caianiello",}}\\
 {\em {\small Universit\`{a} di Salerno, I-84081 Baronissi, Salerno,}}\\  
 {\em {\small $^{2}$Dipartimento di Scienze Fisiche, Universit\`{a} di Napoli,}}\\ 
 {\em {\small $^{3}$Istituto Nazionale di Fisica Nucleare, Sezione di Napoli,}}\\
 {\em {\small Mostra d'Oltremare pad. 19 I-80125 Napoli,}}\\
 {\em {\small $^{4}$Osservatorio Astronomico di Capodimonte,}}\\
 {\em {\small Via Moiariello, 16 I-80131 Napoli, Italy.}}}
	      \date{}
	      \maketitle
	      \begin{abstract}
In the framework of extended gravity theories, we discuss the meaning of
a time dependent
``cosmological constant" and give a set of conditions
to recover  asymptotic de Sitter behaviour 
for a  class of cosmological models independently of initial data. 
To this purpose we introduce  a time--dependent (effective)
quantity which asymptotically becomes the true cosmological constant.
We will deal with scalar--tensor,
fourth and higher than fourth--order theories.
 
 	     \end{abstract}

\vspace{20. mm}
e-mail address:\\
capozziello @vaxsa.csied.unisa.it\\
deritis@axpna1.na.infn.it\\
marino@cerere.na.astro.it
 	      \vfill
	      \end{titlepage}

\section{\normalsize \bf Introduction}
The determination of cosmological constant has become one of the main
issue of modern physics since by fixing its value, one could contribute to
select self--consistent models  of fundamental physics and cosmology.
Briefly, its determination should provide the  gravity
vacuum state \bib{weinberg}, should make to undersand the mechanism
which led the early universe to the today observed large scale structures
\bib{guth},\bib{linde}, 
and to predict what will be the fate of the whole universe (no--hair
conjecture) \bib{hoyle}.

From the  cosmological point of view, 
the main feature of inflationary models is the presence
of a finite period during which the expansion is de Sitter (or quasi--de 
Sitter or power law): this fact implies that 
the expansion of the scale factor $a(t)$ 
is superluminal (at least $a(t)\sim t$, in general $a(t)\sim \exp H_{0}t$
where $H_{0}$ is the Hubble parameter nearly constant for a finite period) 
with respect to the comoving proper time $t$. 
Such a situation arises in  presence of an 
effective energy--momentum tensor which is approximately proportional 
(for a certain time) to the metric tensor and takes place in various
gravitational theories: \ie the Einstein gravity
minimally coupled with a scalar field \bib{guth},\bib{linde},
fourth or higher--order  gravity
 \bib{starobinsky},\bib{ottewill},\bib{schmidt},\bib{kluske} scalar--tensor 
 gravity \bib{la},\bib{cimento}. 

Using 
conformal transformations (by which higher--order geometric terms and
nonminimally coupled fields are reduced to the Einstein gravity plus 
(non)interacting scalar fields 
\bib{maeda},\bib{rainer},\bib{teyssandier},\bib{wands},\bib{conf}) 
all of these approaches 
can furnish dynamical models where
some scalar fields are displaced from their 
equilibrium states ({\it false vacuum states})
and then  evolve sufficiently slow toward the minima of a potential,
in general toward  new equilibrium states ({\it true vacuum states}). 
If more than one  scalar field
undergo such a phenomenology, one can get multiple inflation 
\bib{schmidt},\bib{polarski}. 

However, in all these schemes, we have to provide the solution of 
the so called "flatness", "monopole" and "horizon" problems \bib{guth},
\bib{linde} and, besides,
 mechanisms able to give a natural explanation of the observed small
inhomogeneities in the large scale structure of the universe 
\bib{brandenberger}.

Several inflationary models are affected by the shortcoming of "fine 
tuning" \bib{albrecht}, that is inflationary phase proceeds from very
special initial conditions, while a natural issue would be to get 
inflationary solutions as attractors for a large set of initial conditions.
Furthermore, the same situation should be achieved also in the future:
if a remnant of cosmological constant is today observed, the universe
should evolve toward a final de Sitter stage. A more precise formulation
of such a conjecture is possible for a restricted class of cosmological 
models, as discussed in \bib{wald}.
We have to note that the conjecture holds when any ordinary matter field,
satisfies the
dominant and strong energy conditions \bib{hawking}. However it is possible
to find models which explicitly violate such conditions but satisfies 
no--hair theorem requests. Precisely, this fact happens if extended 
gravity theories are involved and matter is in the form of scalar fields,
besides the ordinary perfect fluid matter \bib{quartic}.

In any case, we need a time variation of cosmological constant
to get successful inflationary models, to be in agreement with
observations, and to obtain a  de Sitter stage
 toward the future. In other words, this means that cosmological 
constant has to
acquire a great value in early epoch (de Sitter stage), has to undergo a phase
transition with a graceful exit (in order to recover the observed
Friedman stage of present epoch) and has to result
in a small remnant toward the future \bib{lambdastep}. 

In this context, a fundamental issue is to select the classes of 
gravitational theories and the conditions which
"naturally" allow to recover an effective 
time--dependent cosmological constant without considering special
initial data.

This paper is devoted to this problem. We take into consideration
extended gravity theories and try to select conditions to obtain 
effective time--dependent cosmological constant.
 The main request is that such effective
cosmological constants evolve (at least asymptotically) 
toward the actual cosmological 
constant which means that the de Sitter behaviour has to be recovered. 
 
In Sect.2, we discuss the effective cosmological constant and the properties 
of the de Sitter space--times. Sect.3 is devoted to the general discussion of 
 extended gravity theories  involving 
higher--order corrections to the Einstein--Hilbert action and  
scalar--tensor couplings.  
 In Sects.4,5,6,7, we take into account specific realizations of such a 
theories that is
scalar--tensor, fourth--order, fourth--order--scalar tensor and higher
than fourth--order gravity theories, respectively, 
and the conditions to obtain
de Sitter.  Some cosmological models, as examples, 
are outlined in Sect.8.

Discussion and conclusions are drawn in Sect.9.

\section{\normalsize\bf The effective cosmological constant}

Before starting with our analysis, it is worthwhile to spend some words
on what we mean by "effective cosmological constant".
The no--hair conjecture  \bib{hoyle}
 claims that if there is a positive cosmological 
constant, all
the expanding universes will approach a de Sitter behaviour. 
In other words, if a cosmological constant is present,  
 the universe will become  homogeneous and isotropic with any initial 
conditions.
However,  there is no general proof  of such a conjecture and
 there are counter--examples of initially expanding and then
recollapsing universes which never become de Sitter \bib{cotsakis}.

A simplified version of the conjecture can be
proved. It is:

 {\it All Bianchi cosmologies (except IX), in presence of 
a positive cosmological constant, asymptotically approach the de Sitter 
behaviour} \bib{wald}. 

It is worthwhile to note that
here the cosmological constant is a true constant (put by hands)
and the contracted Bianchi 
identity is not used, then the  proof is independent of the
 evolution of  matter. 
In order to extend no--hair conjecture to generalized theories of 
gravitation, we have to introduce different sets of conditions
(respect to those given in \bib{wald}) since the cosmological constant is not
introduced {\it a priori}, but it can be "recovered" from  dynamics
of  scalar fields (considering as a sort of "scalar fields" also 
higher--order geometric terms in the gravitational Lagrangian \bib{schmidt},
\bib{wands}).  Such conditions must not use
 the "energy conditions" \bib{hawking}, 
but they have to allow the introduction of a sort of {\it "effective
cosmological constant"} which asymptotically becomes the de Sitter constant.
This feature is due to the fact that,
in an expanding universe, all the contributions to the energy density
and to the Ricci tensor has to decay as some power of the scale factor $a(t)$.
The cosmological constant is the only term that does not decrease with time.
Hence, in an expanding universe, $\Lambda$ is the asymptotically dominant term
in the Einstein equations
(\ie the $(0,0)$ Einstein equation becomes 
${\displaystyle H^{2}=\frac{\Lambda}{3}}$ with $H$ the Hubble parameter);
giving rise to a de Sitter spacetime.
Actually, the effective cosmological constant is time--dependent but,
at the end, it has to coincide with the de Sitter one (the real gravitational
vacuum state).  Then, given any extended theory of gravity, it could be
possible, in general, to define an effective time varying
 cosmological constant which becomes the "true" cosmological constant
if and only if the model asymptotically approaches de Sitter
(that is only asymptotically no--hair conjecture is recovered). This
fact will introduce constraints on the choice of the gravitational couplings,
scalar field potentials and higher--order geometrical terms which combinations
can be intended as components of the effective stress--energy tensor.   

\section{\normalsize\bf The extended gravity theories and cosmology}

There is no {\it a priori} reason to restrict the gravitational Lagrangian
to a linear function of the Ricci scalar $R$ minimally coupled with
matter \bib{francaviglia}. 
Additionally, we have to note that, recently, some authors have taken
into serious consideration the idea that
there are no "exact" laws of physics but that the Lagrangians of physical
interactions are "stochastic" functions with the property that local
gauge invariances (\ie conservation laws) are well approximated in the low
energy limit  and physical constants can vary \bib{ottewill}. 
This scheme was adopted in order to treat
the quantization on curved spacetimes and the result was that the 
interactions
among quantum scalar fields and background geometry or the gravitational
self--interactions yield corrective terms in the Einstein--Hilbert Lagrangian
\bib{birrell}. Futhermore, it has been realized that 
such corrective terms are inescapable if we want to obtain the effective 
action of quantum gravity on scales closed to the Planck length 
\bib{vilkovisky}. They are higher--order terms in curvature invariants
as $R^{2}$, $R\umunu R\dmunu$, $R^{\mu\nu\alp\beta}R_{\mu\nu\alp\beta}$,  
$R\Box R$, or $R\Box^{k}R$, or nonminimally
coupled terms between scalar fields and geometry as $\p^{2}R$.
Terms of these kinds arise also in the effective Lagrangian
of strings and Kaluza--Klein theories when the mechanism of dimensional
reduction is working \bib{veneziano}.

From a completely different point of view, these alternative theories become
interesting when one try to incorporate the Mach principle in gravity and 
to consider the concept of "inertia" in connection to the various 
formulations of equivalence principle. For example, the Brans--Dicke theory
is a serious attempt of alternative theory to the Einstein gravity: it
takes into consideration
a variable  Newton gravitational constant whose dynamics is governed by
a scalar field nonminimally coupled with geometry. In such a way,
the Mach principle is better implemented 
\bib{cimento},\bib{brans},\bib{sciama}.

Besides fundamental physics motivations, all these theories have acquired a 
huge
interest in cosmology due to the fact that they "naturally" exhibit 
inflationary behaviours and that the related cosmological models seem very
realistic \bib{starobinsky},\bib{la}. Furthermore, it is possible to show
that, via conformal transformations, the higher--order and nonminimally
coupled terms ({\it Jordan frame}) always corresponds to the 
Einstein gravity plus one or more
than one minimally coupled scalar fields ({\it Einstein frame})
\bib{teyssandier},\bib{maeda},\bib{wands},\bib{conf},\bib{gottloeber}. 
More precisely (in the Jordan frame),
the higher--order terms appear always as an enhancement of order two
in the equations of motion. For example,
a term like $R^{2}$ gives fourth order equations \bib{ruzmaikin}, $R\Box R$
gives sixth
order equations \bib{gottloeber},\bib{sixth}, 
$R\Box^{2}R$ gives eighth order equations 
\bib{eight} and so on.
By the conformal transformation, any 2--orders give a scalar field: 
for example, fourth--order gravity gives Einstein plus one scalar field,
sixth order gravity gives Einstein plus two scalar fields and so on 
\bib{schmidt},\bib{gottloeber}. This feature results very interesting if
we want to obtain multiple inflationary events since an early stage
could select ``very'' large scale structures (clusters of galaxies today),
 while a late stage could select ``small'' large scale structures (galaxies 
today) \bib{sixth}. The philosophy is that each inflationary
era is connected with the dynamics of a scalar field \bib{polarski}.
Furthermore, these extended schemes naturally could solve the problem
of "graceful exit" bypassing the shortcomings of former inflationary models
\bib{la},\bib{aclo}.
  
Here we want to consider such theories, in general, and to ask for 
recovering  the de Sitter behaviour in the related cosmological models.

Let us start with the most general class of higher--order--scalar--tensor 
theories
in four dimensions. They can be assigned by the action
\beq
\label{3.1}
{\cal A}=\int d^{4}x\sqrt{-g}\left[F(R,\Box R,\Box^{2}R,..\Box^kR,\p)
 -\frac{\epsilon}{2}
g\umunu \phi\ddemu \phi\ddenu+{\l}_{m}\right],
\eeq
where $F$ is an unspecified function of curvature 
invariants
and of a scalar field $\p$. The term ${\l}_{m}$ is the  minimally 
coupled ordinary matter
contribution. We shall use physical units $8\pi G=c=\hbar=1$;
 $\epsilon$ is a constant which specifies the theory.

The field equations are obtained by varying (\ref{3.1}) with respect to
the metric $\gd$.  We get
\beqa
\label{3.2}
G\umunu&=&\frac{1}{{\cal G}}\left[T\umunu+\frac{1}{2}\gu (F-{\cal G}R)+
(g^{\mu\lambda}g^{\nu\sigma}-\gu g^{\lambda\sigma}) 
{\cal G}_{;\lambda\sigma}\right.\nonumber\\
& & +\frac{1}{2}\sum_{i=1}^{k}\sum_{j=1}^{i}(\gu g^{\lambda\sigma}+
  g^{\mu\lambda} g^{\nu\sigma})(\Box^{j-i})_{;\sigma}
\left(\Box^{i-j}\frac{\pa F}{\pa \Box^{i}R}\right)_{;\lambda}\nonumber\\
& &\left.-\gu g^{\lambda\sigma}\left((\Box^{j-1}R)_{;\sigma}
\Box^{i-j}\frac{\pa F}{\pa \Box^{i}R}\right)_{;\lambda}\right]\,,
\eeqa
where
\beq
\label{3.3}
G\umunu=R\umunu-\frac{1}{2}\gu R
\eeq
is the Einstein tensor and
\beq
\label{3.4}
  {\cal G}\equiv\sum_{j=0}^{n}\Box^{j}\left(\frac{\pa F}{\pa \Box^{j} R}
\right)\;.
\eeq
The differential Eqs.(\ref{3.2}) are of order $(2k+4)$.
The stress--energy tensor is due to the kinetic part of the scalar field and to
the ordinary matter:
\beq
\label{3.5}
T\dmunu=T^{(m)}\dmunu+\frac{\epsilon}{2}[\p\ddemu\p\ddenu-
\frac{1}{2}\p\udea\p\ddea]\;.
\eeq
The (eventual) contribution of a potential $\v$ is contained in the 
definition of $F$. From now on, we shall indicate by a capital $F$
a Lagrangian density containing also the contribution of a potential $\v$ 
and by  $f(\p)$, $f(R)$, or $f(R,\Box R)$ a function of such fields
 without potential.

By varying with respect to the scalar field $\p$, we obtain the Klein--Gordon
equation 
\beq
\label{3.6}
\epsilon\Box\p=-\frac{\pa F}{\pa\p}\,.
\eeq
Several approaches can be used to treat such equations. For example,
as we said, by a conformal transformation, 
it is possible to reduce an extended theory of gravity
to a (multi)scalar--tensor theory of gravity 
\bib{schmidt},\bib{wands},\bib{gottloeber},\bib{damour}.
Here we want to discuss under what conditions  it is
possible to obtain asymptotic de Sitter behaviour from (\ref{3.2})
considering some cases of physical interest.
Our discussion will be in Jordan frame. For a detailed  exposition of the 
differences between the Jordan and the Einstein frames, see \eg 
\bib{conf},\bib{magnano}: 
the debate of which of them is the true physical frame is still open.

\section{\normalsize\bf Scalar--tensor gravity}
The scheme which we adopt to find the conditions  
for an asymptotic no--hair theorem
is outlined, for scalar--tensor gravity, in \bib{lambdat} and in 
\bib{bianchi}. Here, for the sake of completeness, we shall carry
the same discussion and enlarge it to other extended gravity theories.

With the choice
\beq
\label{4.1}
F=\f R-\v\;,\;\;\;\;\;\;\epsilon=-1\;,
\eeq
we recover the  scalar--tensor gravity in which a scalar field
$\p$ is nonminimally coupled with the Ricci scalar \bib{cimento},\bib{nmc}.
Here, we do not fix the coupling $\f$ and the potential $\v$ but we
ask for recovering (in general) the de Sitter behaviour by restoring
the cosmic no--hair theorem \bib{lambdat}. As we shall see, this request
will fix a class of couplings and potentials. 

The action (\ref{3.1}) now becomes 
\beq
\label{4.8}
{\cal A}=\int d^{4}x\sqrt{-g}\left[\f R +\frac{1}{2}
g\umunu \phi\ddemu \phi\ddenu-V(\phi)+{\l}_{m}\right]\;,
\eeq
while the Einstein equations are
\beq
\label{4.9}
G\dmunu=\tilde{T}\dmunu=-\frac{1}{2\f}T\dmunu\,,
\eeq
with stress--energy tensor defined as
\beq
\label{4.10}
T\dmunu=T^{(\p)}\dmunu+T^{(m)}\dmunu\;;
\eeq
and the scalar field part
\beq
\label{4.11}
T^{(\p)}\dmunu=
\phi\ddemu \phi\ddenu-\frac{1}{2} g\dmunu \phi\ddea \phi\udea+g\dmunu\v
+2g\dmunu\Box\f-2\f\ddemunu\,,
\eeq
in which we have assembled also the terms coming from the coupling $\f$
which were outside $T\dmunu$ in (\ref{3.2}). Here ${\g}=\f$.
The standard Newton gravitational constant is replaced by the effective 
coupling
\beq
\label{4.12}
G_{eff}=-\frac{1}{2\f}\,.
\eeq
Einstein gravity is restored when  $\f$ assumes the value $-1/2$.

The Klein--Gordon  equation is 
\beq
\label{4.13}
\Box \p-R\fp+\vp=0\,,
\eeq
where the prime means the derivative with respect to $\p$. 
The derivation of such an equation from the contracted Bianchi identity for
$T\dmunu$  is discussed in  \bib{nmc}.
As a general feature, the models described by (\ref{4.8}) are singularity free 
\bib{quartic}; 
then, there are no restrictions on the interval of time on which
the scale factor $a(t)$ and the scalar field $\p(t)$ are defined.
As we shall see in this context, it is possible to introduce  a sort of time 
dependent (effective) {\it cosmological constant}
and this will be the goal for any extended gravity theory which we shall
take into consideration.

For the sake of simplicity, 
we  develope our considerations in a 
FRW--flat spacetime, but the results can be easily extended
 to any
homogeneous cosmological model including also
Kantowski--Sachs models \bib{lambdat},\bib{bianchi},\bib{ryan}.
To get our goal, we shortly sketch the scheme already presented 
in \bib{lambdat}.

From (\ref{4.9}), 
 using a Friedman--Robertson--Walker (FRW) flat metric
\beq
\label{FRW}
ds^{2}=dt^{2}-a(t)^{2}(dx^{2}+dy^{2}+dz^{2})\,,
\eeq
where  $a=a(t)$ is the scale factor of the universe, 
we get
the (cosmological) Einstein equations
\beq
\label{4.7}
H^{2}+\frac{\dot{f}}{f}H +\frac{\rho_{\p}}{6f}+\frac{\rho_{m}}{6f}=0\,,
\eeq
\beq
\label{4.88}
\dot{H} = -\left(H^2 +\frac{V}{6f}\right) - H\frac{\dot{f}}{2f} +
\frac{\dot{\p}^2}{6f}-
-\frac{1}{2}\frac{\ddot{f}}{f} + \frac{3p_{m}+\rho_{m}}{12 f}.
\eeq
where
\beq
\rho_{\p}=\frac{1}{2}\dot{\p}^{2}+\v\,,
\eeq
$\rho_m$ is the ordinary matter density  and the equation of state
\beq
\label{state}
p_{m}=(\gamma-1)\rho_{m}\,,
\eeq
is assumed.

Eq.(\ref{4.7}) can be rewritten as:
\beq
\label{4.7'}
{\cal P}(H) \equiv 
\left(H-\Lambda_{eff,\,1}\right)\left(H-\Lambda_{eff,\,2}\right)=
-\frac{\rho_{m}}{6f}\,.
\eeq
${\cal P}(H)$ is a second degree polynomial in $H$, and
\beq
\label{4.99}
\Lambda_{eff,\,1,2}=-\frac{\dot{f}}{2f}\pm
\sqrt{\left(\frac{\dot{f}}{2f}\right)^{2}-\frac{\rho_{\p}}{6f}}\,.
\eeq
$\Lambda_{eff,\,1,2}$ have to be real.
Let us  assume, for large $t$,
\beq
\label{i}
\frac{\dot{f}}{f}\longrightarrow \Sigma_{0}\;,
\;\;\;\;\;\;
\frac{\rho_{\p}}{6\f}\longrightarrow\Sigma_{1}\;;
\eeq
where $\Sigma_{0,1}$ are two constants depending on the parameters
present in the  coupling and  the potential. 
From these two hypotheses,
 $\Lambda_{eff,\,1,2}$ asymptotically go to constants.
{\it Vice versa}, if for large $t$,
$\Lambda_{eff,\,i}\rightarrow \Lambda_{i}$, where $\Lambda_{i}$ are constants, 
 $\dot{f}/f$ and $\rho_{\p}/6f$ become
constants.
Then hypotheses (\ref{i})  
are necessary and sufficient conditions
since  $\Lambda_{eff,\,1,2}$ are asymptotically constants.
  
If,  asymptotically, the sign of $\f$ is constant 
(this is a natural assumption), 
 we have  two cases: $f(t\gg 0)\leq 0$ and   $f(t\gg 0)\geq 0$. Then
being also $\dot{f}/f$  asymptotically constant,
 each of the above cases has two subcases  related
to the sign of $\dot{f}$. 

The case
$f(t\gg 0)\leq 0$ is physically relevant while 
the other one (repulsive gravity) tells us that 
recovering a de Sitter asymptotic 
behaviour for $a(t)$ is not  connected to  the  sign of 
gravity. 

Let us  now consider the case $f(t\gg 0)\leq 0$ and $\dot{f}(t\gg 0)\leq 0$:
from  (\ref{i}) we have $\Sigma_{0}\geq 0$. 
Eq.(\ref{4.7'}) gives
\beq
\label{4.7'''}
{\cal P}(H)\geq 0,
\eeq
then we have $H \geq \Lambda_{1}$, $H \leq \Lambda_{2}$.
For the two $\Lambda_{i}$, we obtain the asymptotic expressions:
\beq
\Lambda_{1,2} = -\frac{\Sigma_{0}}{2} \pm 
\sqrt{\left(\frac{\Sigma_{0}}{2}\right)^2 +|\Sigma_{1}|} \,,
\eeq
Considering Eq.(\ref{4.88}), 
if 
\beq
\label{iii}
H^2 \geq \frac{V}{6|f|}\,\,,
\eeq
we obtain 
\beq
\label{4.8'''''}
\dot{H} \leq 0\,.
\eeq
In other words, from the two disequalities on ${\cal{P}}(H)$
and on $\dot{H}$ we find that $H(t)$ has  a horizontal asymptote,
or, equivalently,  $H$ goes to a constant (see \bib{lambdat}). 
Then the universe, for large $t$, has a de Sitter behaviour,
(\ie $a(t)\sim \exp (\alp t)$,  where $\alp$ is a constant).
Due to the  conditions (\ref{i}), the constant 
asymptotic sign of $f(\p(t))$ and  the condition
(\ref{iii}),   the 
universe, for large $t$, expands as de Sitter, even if  it is not fixed
the parameter which specifies such an expansion, i.e. 
the effective cosmological constant.         
If we compare the conditions in \bib{wald} with ours, we have:
\beqa
\mbox{(Wald's conditions)}\;\;\;\; & &\;\;\;\mbox{(our asymptotic conditions)}\nonumber\\
i)\;\;\left(H-\sqrt{\frac{\Lambda}{3}}\right)\left(H
+\sqrt{\frac{\Lambda}{3}}\right)\geq 0\;\;\;\;\;\;\; 
&\;\;\;\Longleftrightarrow\;\;\;&\;\;\;\;\;A)\;\;(H-{\Lambda}_{1})
(H-{\Lambda}_{2})\geq 0,\nonumber\\
ii)\;\;\;\;\;\;\;\;\;\;\;\;\;\;\;\;\dot{H}\leq\frac{\Lambda}{3}-H^{2}\leq 0 
&\;\;\;\;\;\Longleftarrow\;\;\;&\;\;\;\;\;B)\;\; \dot{H}\leq 0\,.\nonumber
\eeqa
The hypothesis (\ref{iii}), when $\p \rightarrow $constant is nothing else but 
${\displaystyle H^2 \geq \frac{\Lambda}{3}}$ 
(in our unit $G_{eff} \rightarrow G_{N}$
if ${\displaystyle \f \rightarrow - \frac{1}{2}}$); 
that is we recover the standard
case where ${\displaystyle \frac{V}{6|f|}=const}$ can be interpreted as 
the cosmological constant. 
By some algebra, it is easy to show that such a hypothesis is equivalent to 
\beq
\label{pippo}
\left(\frac{1}{12f}\right)
\frac{{\dot \p}^2}{2V} \geq \left(\frac{f'}{f}\right)^{2} =
 \left(\frac{G'_{eff}}{G_{eff}}\right)^2 \,,
\eeq
where $G'_{eff}=dG_{eff}/d\phi$.
That is we have a constraint on the minimal value
of the (effective) 
ratio of the  kinetic energy  
and the potential energy of the scalar field given by $G_{eff}$ and its
derivative $G'_{eff}$.

Since $a(t)$ behaves like de Sitter for large $t$,
we have to  see if it is possible to fix $\alp$ in order 
to recover  the "true"
cosmological constant.
 To this purpose,
the Bianchi contracted identity for  matter is 
needed (we have not used any Bianchi
 identity to find  the asymptotic behaviour of $a(t)$).
 We get   $\rho_{m} = Da^{-3\gam}$
(by the state equation 
$p_{m} =(\gam -1)\rho_{m}$ with $1\leq \gam \leq 2$; $D$ is  an 
integration constant). 
Introducing this result in Eq.(\ref{4.7}),
for large $t$, we have 
\beq
(H - \Lambda_{1})(H + |\Lambda_{2}|) = 
\frac{D}{|f_{0}|}e^{-(3\gam\alp + \Sigma_{0})t}\,,
\eeq
being $3\gam \alp+\Sigma_{0} \geq 0$.
Then we get
$(H - \Lambda_{1})(H + |\Lambda_{2}|)\rightarrow 0$,
i.e. $H \rightarrow \Lambda_{1}$.
The (effective) matter content, 
$\rho_{m}/6\f$,
tells us how much $H$ is "distant" from the true de Sitter behaviour 
given by
the cosmological constant $\Lambda_{1}$. In other words, we do not use 
the Bianchi identity for finding the type of expansion, 
we only use it to select (asymptotically) the specific value of 
the "cosmological constant". In any case, we have to note that, for
$\rho_{m}=0$, $H=\Lambda_{eff,\,1}$ is a solution for any $t$.
Actually the effective cosmological constant that we have obtained
$\it {via}$ such a procedure will depend on the parameters of
the effective gravitational coupling $f(\p)$  and  
 the potential $V(\p)$.

In a certain sense,
the  approach followed in \bib{wald} is reversed: 
there, $\Lambda$ (constant) is
introduced {\it a--priori} 
and this leads, under certain hypotheses,  to a de Sitter expansion.
Here, the de Sitter expansion is recovered under  different  
hypotheses, and this 
(together with the contracted Bianchi identity  for matter)
selects the effective cosmological constant.  
Moreover, we have 
obtained such a result without assuming to recover the standard
gravity (i.e. we do not need that $G_{eff}\rightarrow G_{N}$).
If we now consider also the Klein--Gordon equation, from the 
conditions (\ref{i}), 
we get, for large $t$, 
that  $\dot{\p}^{2}/\f$ goes to a constant. Being $f(\p(t\gg 0))\leq 0$,
such a constant has to be negative: this request implies 
$|\Sigma_{1}| \geq 2\Sigma_{0}^{2}\,$ \bib{lambdat}.
By this last condition and  (\ref{i}), we get also
that  the potential has to be (asymptotically) non--negative.
In the case $\Sigma_{0} = 0$,
we get that only ${\displaystyle\frac{V}{6f}}$ is different from zero,
giving rise to the
expression ${\displaystyle\frac{V}{6f}}(t\gg 0) = -\Sigma_{1}^2$ which 
identifies the cosmological (asymptotic) constant \bib{lambdat}.

Let us now consider  the   case
$f(\p(t\gg 0))\leq 0$, that is $\dot f(\p(t \gg 0))\geq 0\,.$
Here $\Sigma_{0} \leq 0$ while everything else is the same as 
above.
In particular, the signs of the asymptotic values of 
$\Lambda_{1,2}$ are the same.
From the compatibility of all the  hypotheses  
 we  made with the Klein--Gordon equation 
 we get $\dot{\p}^2/\f\geq 0$,  being $\Sigma_{0}\leq 0$.
Then the compatibility
between (\ref{i}) and the Klein--Gordon equation implies, 
for large $t$, that the scalar
field has to go to a constant. In our units, $f\rightarrow -1/2$,
and $\Lambda \rightarrow \sqrt{V(t\gg 0)/3}$.

Finally, let us consider the case of asymptotically repulsive 
gravity, that is
$f(\p(t\gg 0))\geq 0$.
Also here we have two subcases, $\dot{f}(\p(t \gg 0))\leq 0$ and
$\dot{f}(\p(t \gg 0))\geq 0$. 
This unphysical situation 
tells us that the (asymptotic) de Sitter
behaviour and the recovering of  
standard (attractive) gravity are not necessarily related.
Of course, the condition on the reality of $\Lambda_{i}$  has to be
carefully considered.
The most interesting subcase is $\dot{f}\leq 0$. Here, we have 
two (asymptotic)
positive cosmological constants, that is  
$\Lambda_{eff\,1,2}\rightarrow \Lambda_{1,2}\geq 0,$
$\Lambda_{1}\geq \Lambda_{2}$.
Being $-\rho_{m}/6f\leq 0$,
we have $\Lambda_{1}\leq H\leq \Lambda_{2}$. Then, it is crucial 
to know the sign of $\dot H$: if $\dot H \geq 0$ the effective $\Lambda$
is given by the $max\,(\Lambda_{1},\Lambda_{2})$; viceversa,
if $\dot H \leq 0$,  $\Lambda$ is given by the 
minimun between them.

In conclusion, in scalar--tensor theories, it is possible to extend
asymptotically the no--hair theorem if an effective cosmological constant
is introduced and, asymptotically, it becomes the true cosmological constant.
Starting from these results, we enlarge the discussion to fourth--order,
fourth--order--scalar tensor, and higher than fourth--order theories by 
applying the same scheme.  

\section{\normalsize\bf Fourth-order gravity}

The approach we are discussing works also if 
 the gravitational Lagrangian is nonlinear in
the Ricci scalar (and, in general, in
the curvature invariants).
In this case,  dynamics, (\ie the Einstein equations),
is  of order higher than second (for this reason such theories
 are often called {\it higher--order gravitational theories}). 
 Physically, they are interesting since higher--order terms in curvature
 invariants appear when one performs a one--loop renormalization of
 matter and gravitational fields in curved  background
 (see for example \bib{birrell},\bib{barth}).

In cosmology, such theories can furnish inflationary behaviours 
(see \eg 
\bib{starobinsky}, \bib{muller}, \bib{coa},\bib{mijic}) 
but the usual 
inflaton $\p$ has to be replaced
by its geometric counterpart, the Ricci scalar $R$, called {\it scalaron}.

As we have discussed in Sect. 2,  higher--order theories can be reduced
to  minimally coupled scalar--tensor ones, and {\it vice--versa}, by a conformal
transformation \bib{maeda} so that it is reasonable
that the approach we are dealing with can work in such a context.
Here, we take into account the simplest case, a function $f(R)$.

Let us start from the action
\beq
\label{3.7.1}
{\cal A}=\vol \left[f(R)+{\l}_{m}\right]\,,
\eeq
where, as usual, $R$ is the Ricci scalar. 
It is recovered from the extended action (\ref{3.1}) with the choice
\beq
F=f(R)\,,\;\;\;\;\;\; \epsilon=0\,.
\eeq
By varying Eq.(\ref{3.7.1}), we obtain the field equations
\beq
\label{3.7.2}
f'(R)R\dab-\frac{1}{2}f(R)g\dab=f'(R)\udemunu\left(g_{\alpha\mu}g_{\beta\nu}-
g\dab\gd\right)+T^{(m)}_{\mu\nu}\,,
\eeq
which are fourth order equations, due to $f'(R)\udemunu$. 
The prime indicates now the derivative with respect to $R$
(standard Einstein vacuum equations are immediately recovered if $f(R)=R$).
Eq.(\ref{3.7.2}) can be written in the above Einstein form
${\displaystyle G\dmunu=\tilde{T}\dmunu}$ by defining
\beq
\label{stress}
\tilde{T}\dmunu=\frac{1}{f'(R)}\left\{\half \gd\left[f(R)-Rf'(R)\right]+
f'(R)_{;\mu\nu}-\gd\Box f'(R)+T^{(m)}_{\mu\nu}\right\}\,.
\eeq

The standard (minimally coupled) matter has the same role 
discussed above, \ie it gives
no contribution to  dynamics when we consider the asymptotic behaviour
of  system and, eventually, tells us how much $H$ is "distant" from the
exact de Sitter behaviour. For the sake of simplicity, we discard its
contribution (\ie ${\l}_{m}=0$) 
from now on, taking in mind, however, the previous discussion.

As before, we adopt a FRW metric considering that the results can be
 extended to any Bianchi model. 

What we want show is that there exists a formal analogy (without performing
conformal transformations) between a scalar--tensor theory and a 
fourth--order theory which allows us to use the same above conditions
in order to recover the de Sitter behaviour.

In a FRW metric, the action (\ref{3.7.1}) can be written as
\beq
\label{3.7.3}
{\cal A}=\int {\l}(a,\dot{a},R,\dot{R})dt\,,
\eeq
considering $a$ and $R$ as canonical variables. 
Such a  position seems
arbitrary, since $R$ is not independent of $a$ and $\dot{a}$,
but it is generally used in canonical quantization of higher order 
gravitational theories
 \bib{schmidt},\bib{cimento},\bib{vilenkin}.
In practice, the definition of $R$ by $\ddot{a},\dot{a}$ and $a$
introduces a constraint which
 eliminates the second and higher order derivatives in
 (\ref{3.7.3}),  then this last one produces
a system of second order differential equations in
$\{a,R\}$.
In fact, using a Lagrange multiplier $\lambda$, we have
that the action can be written as
\beq
\label{3.7.4}
{\cal A}=2\pi^{2}\int dt\left\{f(R)a^{3}-\lambda\left[
R+6\left(\frac{\ddot{a}}{a}+\frac{\dot{a}^{2}}{a^{2}}+\frac{k}{a^{2}}\right)
\right]\right\}\,.
\eeq
In order to determine $\lambda$, we have to vary the action with respect 
to $R$, that is
\beq
\label{3.7.5}
a^{3}\frac{df(R)}{dR}\delta R -\lambda\delta R=0\,,
\eeq
from which
\beq
\label{3.7.6}
\lambda =a^{3}f'(R)\,.
\eeq
Substituting into (\ref{3.7.4}) and integrating by parts, we obtain the
Lagrangian \bib{cimento}
\beq
\label{3.7.7}
{\l}=a^{3}\left[f(R)-Rf'(R)\right]+6\dot{a}^{2}af'(R)+
6a^{2}\dot{a}\dot{R}f''(R)-6akf'(R)\,.
\eeq
Then the equations of motion are
\beq
\label{3.7.7'}
\left(\frac{\ddot{a}}{a}\right)f'(R)+
2\left(\frac{\dot{a}}{a}\right)f''(R)\dot{R}+f''(R)\ddot{R}+f'''(R)\dot{R}^{2}-
\frac{1}{2}[Rf'(R)+f(R)]=0\,,
\eeq
and
\beq
\label{3.7.8}
R=-6\left(\frac{\ddot{a}}{a}+\frac{\dot{a}^{2}}{a^{2}}+\frac{k}{a^{2}}\right)\,.
\eeq
The $(0,0)$--Einstein equation, impliying the energy condition $E_{\l}=0$, is
\beq
\label{3.7.9}
6\dot{a}^{2}af'(R)-a^{3}\left[f(R)-Rf'(R)\right]+
6a^{2}\dot{a}\dot{R}f''(R)+6akf'(R)=0\,.
\eeq
Let us now define the auxiliary field
\beq
p\equiv f'(R)\;,
\eeq
so that the Lagrangian (\ref{3.7.7}) can be recast in the form
\beq
\label{3.7.10}
{\l}=6a\dot{a}^{2}p+6a^2\dot{a}\dot{p}-6akp-a^3W(p)\,,
\eeq
where 
\beq
\label{3.7.11}
W(p)=h(p)p-r(p)\,,  
\eeq
with 
\beq
\label{3.7.12}
r(p)=\int f'(R)dR=\int p dR=f(R)\,,\;\;\;\;\;\;
h(p)=R\,,
\eeq
such that $h=(f')^{-1}$ is the inverse function of $f'$.

Considering the FRW  pointlike Lagrangian derived from the action 
(\ref{4.8}), we have 
\bib{cimento},\bib{nmc}
\beq
\label{3.7.13}
{\l}=6a\ad^2\f+6\ad a^2\dot{f}(\p)-6ak\f+a^3\left[\frac{1}{2}\dot{\p}^{2}-\v
\right]\;;
\eeq
so that we get the formal analogy between a fourth--order pointlike Lagrangian
and a nonminimally coupled pointlike Lagrangian in FRW
spacetime. The only difference is that in
fourth-order Lagrangian there is no kinetic term, as ${\displaystyle 
\frac{1}{2}\dot{\p}^2}$, for the field $p$. In this sense, the above 
considerations, which hold for nonminimally coupled theories, 
work also in fourth--order
gravity. A Lagrangian like (\ref{3.7.10}) is a special kind of the so called 
Helmholtz Lagrangian \bib{magnano}.

Dynamical system (\ref{3.7.7'})--(\ref{3.7.9}) becomes
\beq
\label{3.7.14}
6\left[\left(\frac{\add}{a}\right)+\left(\frac{\ad}{a}\right)^{2}
+\frac{k}{a^2}\right]=-\frac{d W(p)}{dp}\,,
\eeq
\beq
\label{3.7.15}
\ddot{p}+2\left(\frac{\ad}{a}\right)\dot{p}+
\left[\left(\frac{\ad}{a}\right)^2+2\left(\frac{\add}{a}\right)+
\frac{k}{a^2}\right]p=-\frac{1}{2}W(p)\,,
\eeq
\beq
\label{3.7.16}
6\left(\frac{\ad}{a}\right)^2 p+6\left(\frac{\ad}{a}\right)\dot{p}+
\frac{6k}{a^2}p=-W(p)\,.
\eeq
We want, also in this case, to obtain an effective cosmological constant.
For semplicity, let us assume $k=0$. Eq.(\ref{3.7.16}) becomes
\beq
\label{3.7.17}
H^2+\left(\frac{\dot{p}}{p}\right)H+\frac{W(p)}{6p}=0\,,
\eeq
which can be recast, as above,
\beq
\label{3.7.18}
(H-\Lambda_{eff,\,1})(H-\Lambda_{eff,\,2})=0\,.
\eeq
Note that now $\rho_{m}=0$, but we can easily consider theories
with $\rho_{m}\neq 0$. The results are the same of previous section.
The effective cosmological constant can be formally defined as
\beq
\label{3.7.19}
\Lambda_{eff\, 1,2}=-\frac{\dot{p}}{2p}
\pm\sqrt{\left(\frac{\dot{p}}{2p}\right)^2-\frac{W(p)}{6p}}\,.
\eeq
We have to note that Eq.(\ref{3.7.18}) defines the exact solutions
$H(t)=\Lambda_{eff,\,1,2}$ which, respectively, separate the region with
expanding universes $(H>0)$ from the region with contracting universes
$(H<0)$. See the discussion in previous section with $\rho_{m}\neq 0$.

In order to restore the asymptotic de Sitter behaviour, we 
rewrite Eq.(\ref{3.7.15}), by using (\ref{3.7.16}), as
\beq
\label{3.7.20}
\dot{H}=-\frac{1}{2}\left(H^2+\frac{W(p)}{6p}\right)-
\frac{1}{2}\left(\frac{\dot{p}}{p}\right)^2 -
\half\frac{d}{dt}\left(\frac{\dot{p}}{p}\right)\,.
\eeq
The effective $\Lambda_{eff,1,2}$ becomes an asymptotic constant for
$t\rightarrow \infty$, if the conditions
\beq
\label{3.7.21}
\frac{\dot{p}}{p}\longrightarrow\Sigma_{0}\,,\;\;\;\;
\frac{W(p)}{6p}\longrightarrow\Sigma_{1}\,,
\eeq
hold. From (\ref{3.7.20}), we get $\dot{H}\leq 0$ if
\beq
\label{3.7.22}
H^{2}\geq -\frac{W(p)}{6p}\,.
\eeq
$\Lambda$, obviously, is real if
\beq
\label{3.7.23}
\left(\frac{\dot{p}}{2p}\right)^2\geq \frac{W(p)}{6p}\,.
\eeq
Conditions (\ref{3.7.21}) gives the asymptotic behaviour of field $p$
and potential $W(p)$.  By a little algebra, we  obtain that
asymptotically must be
\beq
\label{3.7.24}
\Sigma_{0}=0\,,\;\;\;\;\;f(R)=f_{0}(R+6\Sigma_{1})\;;
\eeq
where $f_{0}$ is an arbitrary constant. The asymptotic solution is then
\beq
\label{3.7.25}
H^2=\Sigma_{1}\,,\;\;\;\;\;p=p_{0}\,,\;\;\;\;\;\dot{H}=0\,.
\eeq
From Eq.(\ref{3.7.14}), 
or, equivalently, from the constraint (\ref{3.7.8}), we get
\beq
\label{3.7.26}
R=-12H^2=-12\Sigma_{1}\,.
\eeq
Also here the no--hair theorem is restored without using Bianchi 
identities (\ie the Klein--Gordon equation). The de Sitter
solution of Einstein gravity is exactly recovered if
\beq
\label{3.7.27}
\Sigma_{1}=\frac{\Lambda}{3}\,.
\eeq
It depends on the free constant $f_{0}$ in (\ref{3.7.24}) which is assigned
by introducing ordinary matter in the theory. 
This means that, asymptotically, 
\beq
\label{desit}
f(R)=f_{0}(R+2\Lambda)\,.
\eeq
The situation is not completely
analogue to the scalar--tensor case since the request that asymptotically
$a(t)\rightarrow\exp(\Lambda t)$, univocally "fixes" the asymptotic 
form of $f(R)$. Inversely, any fourth order theory which asymptotically
has de Sitter solutions, has to assume the form (\ref{desit}).

We have to stress the fact that it is
the {\it a priori} freedom in choosing $f(R)$ which allows to recover an
asymptotic cosmological constant (which is not  present in the
trivial case $f(R)=R$, unless it is put by hand) so that de Sitter
solution is, in some sense, intrinsic in higher--order theories
\bib{ottewill},\bib{muller}.

\section{\normalsize\bf Fourth--order--scalar--tensor gravity}

Several effective actions of fundamental physics imply higher--order
geometric terms nonminimally
coupled with scalar fields 
\bib{wands},\bib{veneziano},\bib{vilkovisky},\bib{berkin}.
Such theories have cosmological
realizations which, sometimes, allow to bypass the shortcomings of inflationary
models as that connected with the "graceful exit" and bubble nucleation
(see for example \bib{aclo}). Then it is interesting to ask for the
recovering of de Sitter asymptotic behaviour also for these theories.

With the choice
\beq
\label{5.1}
F=F(R,\p)\,,\;\;\;\;\;\mbox{any}\;\epsilon\,,\;\;\;\;{\l}_{m}=0\,,
\eeq
we obtain  the action

\beq
\label{5.2}
{\cal A}=\int d^{4}x\sqrt{-g}\left[F(R,\p)
 -\frac{\epsilon}{2}
g\umunu \phi\ddemu \phi\ddenu\right]\,,
\eeq
which was extensively studied in \bib{maeda}.

We have put ${\l}_{m}=0$ for simplicity as above. 
Also here, the considerations of
Sect.4 hold. 

The Einstein equations are
\beq
\label{5.3}
G\dmunu=\frac{1}{{\cal G}}\left\{T\umunu+\frac{1}{2}\gd(F-{\cal G}R)+
\left[{\cal G}\ddemunu-\gd\Box {\cal G}\right]\right\}\,,
\eeq
where
\beq
\label{5.4}
  {\cal G}\equiv\frac{\pa F}{\pa R}\;.
\eeq
and $T\dmunu$ is just the expression 
\beq
\label{5.5}
T\dmunu=\frac{\epsilon}{2}[\p\ddemu\p\ddenu-
\frac{1}{2}\p\udea\p\ddea]\;.
\eeq
The (eventual) contribution of a potential $\v$ is contained in the 
definition of $F$.
By varying with respect to the scalar field $\p$ we obtain the Klein--Gordon
equation of the form (\ref{3.6}).

A pointlike FRW Lagrangian can be recovered by the  technique already used.
In fact, using the Lagrange multiplier $\lambda$, we have
\beq
\label{5.7}
{\cal A}=2\pi^{2}\int dt\left\{F(R,\p)a^{3}-
\frac{\epsilon}{2}a^{3}\dot{\p}^2-\lambda\left[
R+6\left(\frac{\ddot{a}}{a}+\frac{\dot{a}^{2}}{a^{2}}+\frac{k}{a^{2}}\right)
\right]\right\}\,.
\eeq
In order to determine $\lambda$, we have to vary the action with respect 
to $R$, that is
\beq
\label{5.8}
a^{3}\frac{\pa F(R,\p)}{\pa R}\delta R -\lambda\delta R=0\,,
\eeq
from which
\beq
\label{5.9}
\lambda =a^{3}\frac{\pa F(R,\p)}{\pa R}\,.
\eeq
Substituting into (\ref{5.7}) and integrating by parts, we obtain 
$$
{\l}=a^{3}\left[F(R,\p)-R\frac{\pa F(R,\p)}{\pa R}\right]+
6\dot{a}^{2}a\frac{\pa F(R,\p)}{\pa R}+
6a^{2}\dot{a}\dot{R}\frac{\pa^{2} F(R,\p)}{\pa R^2}+
$$
\beq
\label{5.10}
+6a^{2}\dot{a}\dot{\p}\frac{\pa^{2} F(R,\p)}{\pa \p^2}
-6ak\frac{\pa F(R,\p)}{\pa R}-\frac{\epsilon}{2}a^{3}\dot{\p}^2\,.
\eeq
To get a formal analogy with previous results, we define
\beq
p\equiv\frac{\pa F(R,\p)}{\pa R}\,,
\eeq
and
\beq
\dot{p}=\frac{d}{dt}\frac{\pa F}{\pa R}=\frac{\pa^2 F}{\pa R^2}\dot{R}+
\frac{\pa^{2} F}{\pa R\pa \p}\dot{\p}\,,
\eeq
so that we  have again a Helmholtz point--like Lagrangian.
\beq
\label{5.11}
{\l}=6a\dot{a}^{2}p+6a^2\dot{a}\dot{p}-6akp-\frac{\epsilon}{2}a^3\dot{\p}^2
-a^3W(p,\p)\,,
\eeq
where the potential $W(p,\p)$ corresponds to 
${\displaystyle \left[R\frac{\pa F(R,\p)}{\pa R}-F(R,\p)\right]}$.
Even if (\ref{5.11}) describes the dynamics of  geometry and two scalar fields
$(p,\p)$ it is formally similar to (\ref{3.7.10}) and (\ref{3.7.13})
so that above considerations work also here.
Assuming $k=0$, the  cosmological equations of motion are
\beq
\label{5.13}
H^2+\left(\frac{\dot{p}}{p}\right)H-\frac{\rho}{6p}=0\,,
\eeq
\beq
\label{5.14}
\left[2\dot{H}+3H^2\right]p+\ddot{p}+2H\dot{p}=
-\frac{1}{2}W(p,\p)-\frac{1}{4}\epsilon\dot{\p}^2\,,
\eeq
The Klein--Gordon equations (one for each scalar field) are 
\beq
\label{5.15}
\frac{\pa W(p,\p)}{\pa p}=-6\left(\dot{H}+2H^{2}\right)\,,
\eeq
and
\beq
\label{5.16}
\epsilon [\ddot{\p}+3H\dot{\p}]=\frac{\pa W(p,\p)}{\pa \p}\,.
\eeq
The ``energy--density'' in (\ref{5.13}) depends on two fields and it is
\beq
\label{5.17}
\rho=\frac{\epsilon}{2}\dot{\p}^2-W(p,\p)\,.
\eeq
As usual, we recast Eq.(\ref{5.13}) as
\beq
\left(H-\Lambda_{eff,\,1}\right)\left(H-\Lambda_{eff,\,2}\right)=0\,,
\eeq
and then
\beq
\label{5.18}
\Lambda_{eff,\,1,2}=-\frac{\dot{p}}{2p}\pm
\sqrt{\left(\frac{\dot{p}}{2p}\right)^{2}+\frac{\rho}{6p}}\,.
\eeq
Eq.(\ref{5.14}) can be rewritten as
\beq
\label{5.19}
\dot{H}=-\frac{1}{2}\left(H^2+\frac{W(p,\p)}{6p}\right)-
\frac{1}{2}\left(\frac{\dot{p}}{p}\right)^2 -
\half\frac{d}{dt}\left(\frac{\dot{p}}{p}\right)-
\frac{5}{24}\epsilon\frac{\dot{\p}{^2}}{p}\,.
\eeq
The effective $\Lambda_{eff,1,2}$ become  asymptotically constants for
$t\rightarrow \infty$, if the conditions
\beq
\label{5.20}
\frac{\dot{p}}{p}\longrightarrow\Sigma_{0}\,,\;\;\;\;
\frac{\rho}{6p}\longrightarrow\Sigma_{1}\,,
\eeq
hold. From (\ref{5.19}), we get $\dot{H}\leq 0$ when
\beq
\label{5.21}
H^{2}\geq -\frac{W(p,\p)}{6p}\,,\;\;\;\;\;\;\frac{\epsilon}{p}\geq 0\,.
\eeq
The quantities $\Lambda_{eff,\,1,2}$ converge to  real constants if
\beq
\label{5.22}
\left(\frac{\dot{p}}{2p}\right)^2\geq -\frac{\rho}{6p}\,.
\eeq
In conclusion, the situation is very similar to the fourth--order and 
scalar--tensor cases. However, we have to stress that the quantities 
$W(p,\p)$ and $\Lambda_{eff,\,1,2}$ are functions of two fields and this fact
increase the number of conditions needed to get the 
asymptotic de Sitter behaviour (\eg Eq.(\ref{5.21})). 

\section{\normalsize\bf Higher than fourth--order gravity}

A pure higher than fourth--order gravity theory is recovered, for example,
 with the choice
\beq
\label{6.1}
F=f(R,\Box R)\,,\;\;\;\;\;\epsilon=0\,,\;\;\;\;{\l}_{m}=0\,,
\eeq
which is, in general, an eighth--order theory. If $F$ depends only
linearly on $\Box R$, we have a sixth--order theory. With this
consideration in mind, we shall take into account
the action (\ref{3.1}) which becomes
\beq
\label{6.2}
{\cal A}=\int d^4x\sqrt{-g} f(R,\Box R)\,.
\eeq
The Einstein field equations are now
$$ G\umunu=\frac{1}{\g}\left\{\half g\umunu[f-{\g}R]+{\g}\udemunu-
\gu\Box {\g}-\half\gu[{\fo}_{;\gamma}R^{;\gamma}+{\fo}\Box R]+\right.$$
\beq
\label{6.31}
\left.+\half[{\fo}^{;\mu}R^{;\nu}+{\fo}^{;\nu}R^{;\mu}]\right\}\,,
\eeq
where
\beq
\label{6.4}
{\g}=\frac{\pa f}{\pa R}+\Box\frac{\pa f}{\pa \Box R}\,,\;\;\;\;\;
{\fo}=\frac{\pa f}{\pa \Box R}\,.
\eeq
As above, we can get a FRW pointlike Lagrangian with the position
\beq
{\l}={\l}(a,\ad,R,\dot{R},\Box R,\dot{(\Box R)})\,.
\eeq
Also here, we consider $R$ and $\Box R$ as two independent fields and use the
method of  Lagrange multipliers to eliminate higher derivatives than one 
in time.
The action is
\beq
\label{6.3}
{\cal A}=2\pi^{2}\int dt\left\{f(R,\Box R)a^{3}-
\lambda_{1}\left[R+6\left(\frac{\ddot{a}}{a}+\frac{\dot{a}^{2}}{a^{2}}
+\frac{k}{a^{2}}\right)\right]
-\lambda_{2}\left[\Box R-\ddot{R}-3H\dot{R}\right]\right\}\,.
\eeq
In order to determine $\lambda_{1,2}$, we have to vary the action with respect 
to $R$, and $\Box R$ so that
\beq
\label{6.41}
\lambda_{1} =a^{3}\left[\frac{\pa f}{\pa R}+
\Box\frac{\pa f}{\pa \Box R}\right]\,,
\eeq
\beq
\label{6.5}
\lambda_{2}=a^{3}\frac{\pa f}{\pa \Box R}\,.
\eeq
Substituting into (\ref{6.3}) and integrating by parts, we obtain the
Helmholtz--like Lagrangian
\beq
\label{6.6}
{\l}=a^{3}\left[f-R{\g}+6H^2{\g}+6H\dot{\g}-\frac{6k}{a^2}{\g}-\Box R{\cal F}
-\dot{R}\dot{\cal F}\right]\,.
\eeq
The equations of motion, for $k=0$, are
\beq
\label{6.7}
\dot{H}+\frac{3}{2}H^2+H\left(\frac{\dot{\g}}{\g}\right)+
\frac{1}{2}\left(\frac{\ddot{\g}}{\g}\right)+\frac{\chi}{4{\g}}
+\frac{\dot{R}\dot{\fo}}{2{\g}}=0\;;
\eeq
\beq
\label{6.8}
R=-6[\dot{H}+2H^2]\,,
\eeq
\beq
\label{6.9}
\Box R=\ddot{R}+3H\dot{R}\,,
\eeq
where (\ref{6.8}) and (\ref{6.9}) have the role of  Klein--Gordon
equations for the fields $R$ and $\Box R$ and are also "constraints" for
such fields.
The $(0,0)$--Einstein equation is
\beq
\label{6.10}
H^2+H\left(\frac{\dot{\g}}{\g}\right)+\frac{\chi}{6{\g}}=0\;;
\eeq
while the quantity $\chi$ is defined as
\beq
\label{6.11}
\chi=R{\g}+{\fo}\Box R-f-\dot{R}\dot{\fo}\,.
\eeq
It is interesting to note that $\chi$ has a role similar to that of 
the energy density in previous theories.

As usual, we can define an effective cosmological constant as 
\beq
\label{6.12}
\Lambda_{eff,\,1,2}=-\frac{\dot{\g}}{2{\g}}\pm
\sqrt{\left(\frac{\dot{\g}}{2{\g}}\right)^{2}-\frac{\chi}{6{\g}}}\,.
\eeq
Now, the role of the coupling $f(\p)$ is played by the function
${\g}={\g}(R,\Box R)$.

By substituting Eq.(\ref{6.10}) into Eq.(\ref{6.7}), we get
\beq
\label{6.14}
\dot{H}=-\half\left(H^2-\frac{\chi}{6{\g}}\right)-
\half\left(\frac{\dot{\g}}{\g}\right)^{2}-
\frac{1}{2}\frac{d}{dt}\left(\frac{\dot{\g}}{\g}\right)-
-\frac{\dot{R}\dot{\fo}}{2{\g}}\;.
\eeq
The quantities $\Lambda_{eff,\,1,2}$ becomes asymptotically constants
if
\beq
\label{6.13}
\frac{\dot{\g}}{\g}\longrightarrow\Sigma_{0}\,,\;\;\;\;
\frac{\chi}{6\g}\longrightarrow\Sigma_{1}\,.
\eeq
From (\ref{6.14}), we have $\dot{H}\leq 0$ if
\beq
\label{6.15}
H^{2}\geq 
\frac{\chi}{6{\g}}\,,\;\;\;\;\;\frac{\dot{R}\dot{\fo}}{\g}\geq 0\,.
\eeq
The cosmological constant is real if
\beq
\label{6.16}
\left(\frac{\dot{\g}}{2{\g}}\right)^{2}\geq -\frac{\chi}{6{\g}}\,.
\eeq
This case is analogous to the previous fourth--order--scalar tensor:
There the fields involved where $p,\p$ (or $R,\p$), now they are $R, 
\Box R$. In fact, the quantities $\chi$, ${\g}$, and $\Lambda_{eff}$
are funcions of two fields and the de Sitter asymptotic regime select
particular surfaces $\{R,\Box R\}$.

\section{\normalsize\bf Examples}

The above discussion can be realized on specific cosmological models.
Now, as in \bib{lambdat}, we want to give  examples where, by fixing the 
scalar--tensor or the higher--order theory, the asymptotic de Sitter 
regime is restored in the framework of our generalization of no--hair
theorem.
The presence of standard fluid matter can be implemented by adding the
term
${\displaystyle {\cal L}_{m}=Da^{3(1-\gam)}}$
into the FRW--pointlike Lagrangian \bib{quartic}.
It is a sort of pressure term. We can restrict to the case $\gamma=1$,
(dust) that is ${\l}_{m}=D$, since we are considering asymptotic regimes,
but, in any case the presence of standard fluid matter is not particularly
relevant.

\begin{enumerate}
\item
Let us consider a generic coupling $f(\p)$
and the potential $\v= \Lambda$. Using the  Noether Symmetry Approach
\bib{cimento},\bib{nmc},
we get $\f=\frac{1}{12}\p^{2}+F_{0}'\p+F_{0}$, 
where $F_{0}'$ and $F_{0}$ are two
generic parameters. We have
already discussed such a case in \bib{lambdat} where we show that an 
asymptotic de Sitter regime is restored as soon as $G_{eff}\rightarrow G_{N}$.

\item
In the case 
$\f=k_{0}\p^{2},\;\;\;\;\v=\lambda\p^{2},\;\;\;\;\gam=1\,,$
where $k_{0}<0$ and $\lambda > 0$ are free parameters,
the de Sitter regime is recovered even if solutions do not converge toward 
standard gravity.
The coupling $\f$ is always negative, whereas 
$\v$ is always positive and $\dot f(\p(t \gg 0))<0$ \bib{cimento}.

\item
 Both the above cases can be translated in the fourth--order
formalism and the same results are found if we take into consideration
a theory as $f(R)=R+\alpha R^2$ (see \bib{muller} for the  discussion of the 
case
and \bib{magnano} for the physical equivalence).

\item
The conditions for the existence and stability of de Sitter solutions
for fourth--order theories $f(R)$ are widely discussed in \bib{ottewill}. 
In particular, it is shown that, for $R$ covariantly constant 
(\ie $R=R_{0}$),
as recovered in our case for $R\rightarrow-12\Sigma_{1}$ (see Eq.(\ref{3.7.26}),
the field equations (\ref{3.7.2}) yield the existence condition
\beq
\label{existence}
R_{0}f'(R_{0})=2f(R_{0})\,.
\eeq 
Thus, given any $f(R)$ theory, if there exists a solution $R_{0}$ of
(\ref{existence}) then the theory contains a de Sitter solution.
From our point of view, any time that the ratio  $\dot{f}(R(t))/f(R(t))$ 
converges
to a constant, a de Sitter (asymptotic) solution exists.

On the other hand, given, for example, a theory of the form
\beq
\label{poly}
f(R)=\Sigma_{n=0}^{N}a_{n}R^{n}\,,
\eeq
the condition (\ref{existence}) is satisfied if the polynomial
equation
\beq
\Sigma_{n=0}^{N}(2-n)a_{n}R_{0}^{n}=0\;,
\eeq
has real solutions. Examples of de Sitter asymptotic behaviours
recovered in  this kind of theories are given in \bib{coa}.
\item
Examples of theories higher than fourth--order in which asymptotic de Sitter
solutions are recovered are discussed in \bib{kluske},\bib{eight}.
There is discussed under which circumstances the de Sitter space--time
is an attractor solution in the set of spatially flat FRW models.
Several results are found: for example, a $R^{2}$ non--vanishing term
is necessarily required (\ie a fourth--order term cannot be escaped);
the models are independent of dimensionality of the theory;
more than one inflationary phase can be recovered.

Reversing the argument from our point of view, a wide class of
cosmological models coming from higher--order theories, allows to recover
an asymptotic cosmological constant which seems an intrinsic feature
if Eistein--Hilbert gravitational action is modified by higher--order terms.
In this sense, and with the conditions given above, the cosmological
no--hair theorem is extended.
\end{enumerate}

We conclude the discussion of these examples stressing, again, that
it appears clear that
the (asymptotic) cosmological constant, as introduced in our
approach, depends on the parameters 
appearing into the functions $f(\p)$, $V(\p)$, $f(\p,R)$, or
$f(R,\Box R)$. Furthermore, it depends on the order of higher--order theory
and on the possibility that the condition $\dot{H}\leq 0$ is restored.

\section{\normalsize\bf Discussion and conclusions}

We have discussed the cosmic no--hair theorem in the framework
of extended  theories of gravity by 
introducing a time dependent cosmological "constant".
Such an effective cosmological "constant" has been reconstructed
by  $\dot{G}_{eff}/G_{eff}$ and by $\rho_{\p}/6\f$
but such quantities assume different roles in accordance with the
theory used (higher--order or scalar--tensor).
It is interessing to stress that $R$, $\Box R$ and $\p$ can be all
treated as ``scalar fields"  in the construction of $\Lambda_{eff}$,
\ie all of them give rise to extra--terms in the field equations
which contribute to the construction of an effective stress--energy
tensor $T\dmunu^{(eff)}$.
Actually $\Lambda_{eff}$ has been introduced only in the case 
of homogeneous--isotropic flat cosmologies but it is not difficult to extend
the above considerations to Bianchi models (see \bib{bianchi},\bib{lambdat}).
The  way we have followed to recostruct the no--hair theorem is 
opposite of that  usually adopted: instead of introducing {\it by hands}
a cosmological constant and then searching for the conditions to get an 
asymptotic de Sitter behaviour,
we  find the conditions to get such an  asymptotic behaviour,
and then we  define an effective cosmological "constant"
(actually function of time), which becomes a (true) constant for $t\gg 0$.
Of course, the time behaviour of  $\Lambda_{eff}$ can be  of any
type with respect to the asymptotic constant value \bib{matarrese}.  
Under the hypotheses we  used, the de Sitter asymptotic regime is 
obtained and this is not necessarily connected with  
the recovering of standard Einstein
gravity (which is restored, in our units, for the value 
$\f_{\infty}=-1/2$ of the coupling). 
In other words, the cosmic no--hair theorem
 holds even if we are not in the Einstein regime (it is not even necessary
that the right (attractive gravity) sign of the coupling is recovered). 
Furthermore, the role of the Bianchi contracted
identity for the (standard) matter is to fix (only) 
the specific value of $\Lambda$,
not the kind of the (de Sitter) asymptotic behaviour of $a(t)$.
It is interesting to stress that, by this mechanism, the "amount of $\Lambda$"
is strictly related to the matter content of the universe.
This is worthwhile in connection to the $\Omega$ problem
since it seems that cold dark matter models, with non trivial amount
of cosmological constant, have to be taken into serious consideration for
large scale structure formation \bib{starobinsky1}. 
In conclusion, we want to make two final remarks. The first concerns 
an important question which we have only
mentioned. The way we have followed to introduce the (effective)
cosmological "constant" seems to confine its meaning  only to the cosmological
arena. In the  standard way used to define  such a
quantity, this problem does not exists since it is a true constant of the 
theory and then it is defined independently of any cosmological scenario.
We believe that this question can be solved stressing that cosmology 
has to be taken into account in any other specific physical situation
in relativity. Then  the effective time--dependent
cosmological constant we have introduced gets a role 
of the same kind of the standard $\Lambda$.
From this point of view,  the question we 
are discussing can be answered still using the (standard) way to
define the  cosmological constant, \ie (the cosmological)  $T_{0 0}$. This is
what we actually have done  and what we believe to be
the ingredient to use for understanding the role of  
(effective) cosmological "constant" also in different contexts than cosmology.
Finally, in our construction of $\Lambda$, there is 
a contribution given by the (relative) time variation of the 
effective gravitational coupling: this implies that it would be possible
to compute it, for example, {\it via} the density contrast
parameter. 

A final comment concerns the fact that all extended theories of gravity
can be treated under the same standard of no--hair conjecture.
In this sense, the determination of the effective dynamics of cosmological
constant could be a test on which of them actually works.

\vspace{3. mm}
\begin{centerline}
{\bf Acknowledgments}
\end{centerline}
The authors wish to thank Hans--J\"urgen Schmidt for the useful
suggestions and the careful reading of the manuscript
which allowed  to improve the paper.

\vspace{5. mm}

\begin{centerline}
{\bf REFERENCES}
\end{centerline}
\begin{enumerate}
\item\label{weinberg}
S. Weinberg, \rmp {\bf 61} (1989) 1.
\item \label{guth}
A. Guth, \pr  {\bf D 23} (1981) 347;\\
A. Guth,  \pl  {\bf 108 B} (1982) 389.
\item \label{linde}
A.D. Linde, \pl {\bf B 108}, 389 (1982);\\
A.D. Linde, \pl {\bf B 114}, 431 (1982);\\
A. Linde, \pl {\bf B 129}, 177 (1983);\\
A.D. Linde  \pl {\bf B 238} (1990) 160.
\il{hoyle}
F. Hoyle and J.V. Narlikar, {\it Proc. R. Soc.} {\bf 273A} (1963) 1.
\il{starobinsky}
A.A. Starobinsky, \pl {\bf B 91} (1980) 99.
\il{ottewill}
J. Barrow and A.C. Ottewill, {\it J. Phys. A: Math. Gen.} {\bf 16} (1983)
2757.
\il{schmidt}
H.--J. Schmidt, \cqg {\bf 7} (1990) 1023;\\
H.--J. Schmidt, \pr {\bf D 54} (1996) 7906.
\il{kluske}
S. Kluske and H.--J. Schmidt, {\it Astron. Nachr.} {\bf 317} (1996) 337;\\
S. Kluske in :{\it New Frontiers in gravitation } ed. G. Sardanashvily.
\il{la}
D. La  and P.J. Steinhardt,  {\it Phys. Rev. Lett.} {\bf 62} (1989) 376 \\
D. La,  P.J. Steinhardt  and E.W. Bertschinger,
{\it Phys. Lett.} {\bf B 231} (1989) 231.
\il{cimento}
S. Capozziello, R. de Ritis, C. Rubano, and P. Scudellaro,
{\it La Rivista del Nuovo Cimento} {\bf 4} (1996).
\il{maeda}
K. Maeda, \pr {\bf 39D} (1989) 3159.
\il{rainer}
M. Rainer, \ijmp {\bf 4D} (1995) 397.
\il{teyssandier}
P. Teyssandier and P. Tourrenc, \jmp {\bf 24} (1983) 2793.
\il{wands}
D. Wands, \cqg {\bf 11} (1994) 269.
\il{conf}
S. Capozziello, R. de Ritis, A.A. Marino, \cqg {\bf 14} (1997) 3243.
\il{polarski}
L.A. Kofman, A.D. Linde, A.A. Starobinsky, \pl {\bf 157B} (1985) 361.\\
A. A. Starobinsky, {\it JETP Lett.} {\bf 42} (1985) 152.\\
S. Gottl\"ober, V. M\"uller, and A.A. Starobinsky, \pr {\bf 43D} (1991) 
2510.\\
D. Polarski, A.A. Starobinsky, \np {\bf 385B} (1992) 623.
\il{brandenberger}
R.H. Brandenberger, \rmp {\bf 57}, 1 (1985).\\
V.F. Mukhanov, H.A. Feldman, and  R.H. Brandenberger,
{\it Phys. Rep.} {\bf 215}, 203 (1992).
\item\label{albrecht}
A. Albrecht and P.J. Steinhardt, \prl {\bf 48}, 1220 (1982).
\il{wald}
R.M.Wald, \pr {\bf D28} (1983) 2118.
\il{hawking}
S.W. Hawking and G.F.R. Ellis, {\it The Large--Scale Structure of 
Space--Time}, Cambridge Univ. Press (1973) Cambridge.
\item \label{quartic}
S. Capozziello, R. de Ritis, C. Rubano, and P. Scudellaro,  
\ijmp {\bf D 4} (1995) 767.
\il{lambdastep}
S. Capozziello, R. de Ritis, and A.A. Marino, \ncim {\bf 112 B} (1997) 1351. 
\item\label{cotsakis}
S. Cotsakis and G. Flessas, \pl {\bf B 319} (1993) 69;\\
A.B.Burd and J.D. Barrow,\np {\bf B308} (1988) 929;\\
J.Yokogawa and K.Maeda, \pl {\bf B207} (1988) 31;\\
J.D. Barrow and G. G\"otz, \pl {\bf B 231} (1989) 228.
\item\label{francaviglia}
G. Magnano, M. Ferraris, and M. Francaviglia, {\it Gen. Rel. Grav.}
{\bf 19} (1987) 465\\
M. Ferraris, M. Francaviglia, and I. Volovich,
{\bf Il N. Cim.} {\bf B 108} (1993) 1313.
\item \label{birrell}
A. Zee, {\it Phys. Rev. Lett.} {\bf 42} (1979) 417 \\
L. Smolin,  {\it Nucl. Phys.}  {\bf B 160} (1979) 253 \\
S. Adler, {\it Phys. Rev. Lett.} {\bf 44} (1980) 1567 \\
N.D. Birrell  and P.C.W. Davies  {\it Quantum Fields in Curved Space} (1982)
Cambridge Univ. Press (Cambridge).
\il{vilkovisky}
G. Vilkovisky, \cqg {\bf 9} (1992) 895.
\item \label{veneziano}
M. Green, J. Schwarz and E. Witten, {\it Superstring Theory},
Cambridge Univ. Press, Cambridge (1987).\\
A.A. Tseytlin and C. Vafa \np {\bf B 372} (1992) 443.\\
G. Veneziano \pl {\bf B 265} (1991) 287.\\
M. Gasperini, J. Maharana and G. Veneziano \pl {\bf B 272} (1991) 277.\\
K.A. Meissner and G. Veneziano \pl {\bf B 267} (1991) 33.
\item\label{brans}
C. Brans  and R.H. Dicke,  {\it Phys. Rev} {\bf 124} (1961) 925.
\item\label{sciama}
P.A.M. Dirac, {\it Proc. R. Soc.} London {\bf A 165}, 199 (1937);
D.W. Sciama, {\it Mon. Not. R. Astrom. Soc.} {\bf 113}, 34 (1953);
P. Jordan, {\it Z. Phys.} {\bf 157}, 112 (1959).
\il{gottloeber}
S. Gottl\"ober, H.--J. Schmidt, and A.A. Starobinsky, \cqg {\bf 7} (1990)
893.
\item\label{ruzmaikin}
T.V. Ruzmaikina and A.A. Ruzmaikin, {\it JETP} {\bf 30} (1970) 372.\\
K.S. Stelle, \grg {\bf 9} (1978) 353.\\
R. Schimming and H.--J. Schmidt,
{\it NTM-Schriftenr. Gesch. Naturw.} {\bf 27} (1990) 41. 
\item\label{sixth}
H. Buchdahl, {\it Acta Math.} {\bf 85} (1951) 63.\\
A. Berkin and K. Maeda, \pl {\bf B 245} (1990) 348.\\
L. Amendola, A. Battaglia Mayer, S. Capozziello, S. Gottl\"ober,
V. M\"uller, F. Occhionero, and H.--J. Schmidt, \cqg {\bf 10} (1993) L43.
\item\label{eight}
A. Battaglia Mayer and H.--J. Schmidt, \cqg {\bf 10} (1993) 2441.
\il{aclo}
L. Amendola, S. Capozziello, M. Litterio, and F. Occhionero,
\pr {\bf D 45} (1992) 417.
\item\label{damour}
T. Damour and G. Esposito--Farese, \cqg {\bf 9} (1994) 2093
\il{magnano}
G. Magnano and L.M. Sokolowski, \pr {\bf D 50} (1994) 5039.
\il{lambdat}
S. Capozziello, R. de Ritis, and P. Scudellaro, \pl {\bf A 188} (1994) 130;\\
S. Capozziello, R. de Ritis, C. Rubano, and P. Scudellaro, 
\pl {\bf A 201} (1995) 145;\\
S. Capozziello, R. de Ritis, and A.A. Marino, {\it Helv. Phys . Acta} {\bf 69}
(1996) 241;\\
S. Capozziello and R. de Ritis, \grg {\bf 29} (1997) 1425.
\item\label{bianchi}
S. Capozziello and R. de Ritis, \ijmp {\bf D 5} (1996) 209.
\il{nmc}
S. Capozziello and R. de Ritis,  \cqg {\bf 11} (1994) 107;\\
S. Capozziello and R. de Ritis,  \pl {\bf A 177} (1993) 1;\\
S. Capozziello, R. de Ritis, and P. Scudellaro \ijmp {\bf D 2} (1993) 463;\\
S. Capozziello, R. de Ritis, and C. Rubano, \pl {\bf A 177} (1993) 8;\\
S. Capozziello, M. Demianski, R. de Ritis, and C. Rubano, \pr {\bf D 52} 
(1995) 3288. 
\item\label{ryan}
G.F.R. Ellis and M.A.H. MacCallum, {\it Commun. Math. Phys.} {\bf 12} (1969)
108\\
M.A.H. MacCallum, in {\it General Relativity: An Einstein Centenary Survey},
eds. S.W. Hawking and W. Israel, Cambridge Univ. Press, Cambridge (1979).\\
M.P. Ryan and  L.C. Shepley, {\it Homogeneous Relativistic Cosmologies},
Princeton Univ. Press, Princeton (1975).
\item\label{barth}
N.H. Barth and S.M. Christensen, \pr {\bf 28D} (1983) 1876.
\il{muller}
V. M\"uller and H.--J. Schmidt, {\it Fortschr. Phys.} {\bf 39} (1991) 319.
\il{coa}
S. Capozziello, F. Occhionero, and L. Amendola,
\ijmp {\bf D 1} (1993) 615.
\item\label{mijic}
M.B. Miji\'c, M.S. Morris, and W.M. Suen, \pr {\bf D 34} (1986) 2934
\item\label{vilenkin}
A. Vilenkin, \pr {\bf D 32} (1985) 2511
\il{berkin}
A.L. Berkin and K. Maeda, \pr {\bf 44D} (1991) 1691.  
\item\label{matarrese}
M. Bruni, S. Matarrese, O. Pantano, \prl {\bf 74} (1995) 1916.
\item\label{starobinsky1}
A.A. Starobinsky, in {\it Cosmoparticle Physics 1}, eds. M.Yu. Khlopov
{\it et al.} Edition Frontiers (1996).
\end{enumerate}

\vfill

\end{document}